\documentclass[jkps,preprint,fleqn,showpacs,showkeys]{revtex4}
\usepackage[pdftex]{graphicx}
\usepackage{amssymb}
\usepackage{amsmath}
\usepackage{bm}
\begin{document}
\setcounter{page}{0}
\title[]{Electronic Structure Change of NiS$_{2-x}$Se$_x$ in the Metal-Insulator Transition Probed by X-ray Absorption Spectroscopy}
%\title[]{Metal-Insulator Transition of Ni(S,Se)$_2$ Studied by X-ray Absorption Spectroscopy}
\author{Jinwon \surname{Jeong}}
\author{Kyung Ja \surname{Park}}
\thanks{JJ and KJP equally contributed.}
\author{En-Jin \surname{Cho}}
\author{Han-Jin \surname{Noh}}
\email{ffnhj@jnu.ac.kr}
\thanks{Fax: +82-62-530-3369}
\affiliation{Department of Physics, Chonnam National University, Gwangju 61186}
\author{Sung Baek \surname{Kim}}
\affiliation{The College of Liberal Arts, Konyang University, Chungnam, 32992}
\author{Hyeong-Do \surname{Kim}}
%\affiliation{Department of Physics and Astronomy, Seoul National University, Seoul 151-747, Korea}
\affiliation{Department of Physics and Astronomy, Seoul National University, Center for Correlated Electron Systems, Institute for Basic Science, Seoul 08826}

\date{\today}

\begin{abstract}
The electronic structure change of NiS$_{2-x}$Se$_x$ as a function of Se concentration $x$ has been studied by Ni $L$-edge X-ray absorption spectroscopy (XAS).
The XAS spectra show distinct features in Ni $L_3$ edge, indicating whether the system is insulating or metallic.
These features can be semi-quantitatively explained within the framework of the configurational interaction cluster model (CICM).
In the S-rich region, relatively large charge-transfer energy ($\Delta \sim 5$ eV) from ligand $p$ to Ni 3$d$ states and a little small $p$-$d$ hybridization strength ($V_{pd\sigma} \sim 1.1$ eV) can reproduce the experimental spectra in the CICM calculation, and vice versa in the Se-rich region.
Our analysis result is consistent with the Zaanen-Sawatzky-Allen scheme that the systems in S-rich side ($x \leq$ 0.5) are a charge transfer insulator.
However, it also requires that the $\Delta$ value must change abruptly in spite of the small change of $x$ near $x$=0.5.
As a possible microscopic origin, we propose a percolation scenario where a long range connection of Ni[(S,Se)$_2$]$_6$ octahedra with Se-Se dimers plays a key role to gap closure.

\end{abstract}


% 71.30.+h metal-insulator transitions
% 72.80.Ga	Conductivity of Transition-metal compounds
% 78.70.Dm x-ray absorption spectroscopy


\maketitle

\section{Introduction}

The metal-insulator transition (MIT) of nickel dichalcogenide Ni(S,Se)$_2$ with pyrite structure has long been studied in condensed matter physics\cite{Bouchard, Jarrett, Wilson}.
At the early stage of this material research, NiS$_{2-x}$Se$_x$ was regarded as one of the few ideal examples of band-width controlled MIT by electron correlation effects because this solid-solution does not show a first-order structural distortion that hinders isolating the correlation effects in MIT\cite{Jarrett, Husmann}.
Se substitution ($x$) in S-site of NiS$_2$ is expected to induce (i) an increase of the overlap integrals between Ni 3$d$ and ligand $p$ orbitals and (ii) a decrease of the charge transfer energies from ligand $p$ to Ni 3$d$ states.
In view of the Zaanen-Sawatzky-Allen (ZSA) scheme, the change of the parameter values is to be roughly mapped to a gap closing of the charge transfer insulator NiS$_2$\cite{ZSA}.
This apparently simple issue, however, incorporates several subtle and complicated problems.
Being very different from the transport or optical measurements that indicate 70$\sim$300 meV activation energy in insulating NiS$_2$\cite{Kautz, Mabatah}, the conduction/valence bands photoemission spectra (PES) always show a clear Fermi edge regardless of their Se concentration\cite{Matssura, Mamiya}.
Ni 2$p$ core level PES or X-ray absorption spectroscopy (XAS) studies also have not provided a clear electronic structure change across the MIT.
Further, as is often the case with strongly correlated electron systems, conventional band calculations fail to reproduce the insulating phase\cite{Matssura, Perucchi}.
Due to these difficulties, the microscopic origin of the MIT in NiS$_{2-x}$Se$_x$ has not been well understood in spite of the accumulated data in a long term.

Recently, a few novel scenarios about the microscopic origin of the MIT were proposed based on newly developed methods\cite{ARPES, DMFT, Jeon}.
In theoretical aspects, Kune\v{s} \emph{et al.} exploited the dynamic mean field theory (DMFT) combined with the local density approximation (LDA) to reproduce the gap in NiS$_2$ and addressed that the S-S dimer is the key factor to control the gap size\cite{Kunes}.
Moon \emph{et al.} also employed the DMFT+LDA to reproduce the evolution of the electronic structure of NiS$_{2-x}$Se$_x$, from which they argued that the MIT is a band width-controlled type and that the hybridization strength change between Ni 3$d$ and chalcogen $p$ orbitals is the most important parameter\cite{Moon}.
Experimentally, a soft X-ray angle-resolved photoemission spectroscopic (SX-ARPES) measurement was successfully performed that the spectral weight transfer of the coherent quasiparticle peak to incoherent dispersionless part indicates the increase of the electron correlation that drives the Mott transition\cite{Xu}.

In this paper, we present the electronic structure change of NiS$_{2-x}$Se$_x$ measured by Ni 2$p$ XAS as a function of the Se concentration.
Even though the XAS spectra of the end members, NiS$_2$ and NiSe$_2$, have been reported separately, there is no systematic XAS study for the full series of this system\cite{Charnock, Suga}.
Our measurements revealed that there is a distinct spectral difference between metallic phase and insulating one in the Ni 2$p_{3/2}$ XAS.
We analyzed the spectral change using the configurational interaction cluster model (CICM) calculation with full multiplets and obtained a semi-quantitative agreement with the charge transfer type Mott transition scenario in which the $p$-$d$ charge transfer energy $\Delta$ plays more dominant role in the MIT than the $d$-$p$ hybridization strength, but our result also requires an abrupt change of the $\Delta$ value that cannot be an averaging effect of S and Se electro-negativities.
A possible microscopic origin of the abrupt $\Delta$ change is also discussed in terms of a percolative connection of Ni[(S,Se)$_2$]$_6$ with Se-Se dimers.

\section{Methods}
High quality single crystalline NiS$_{2-x}$Se$_x$ ($x$=0.0, 0.25, 0.3, 0.5, 0.6, and 2.0) were synthesized by the chemical vapor transport method with Cl$_2$ as the transport agent.
Stoichiometric amounts of high purity ($\geq$ 99.99\%) Ni, S, and Se powders were weighed and mixed with a pestle and mortar.
The mixture was loaded in a $\sim$20 cm long quartz ampoule.
The ampoule was evacuated to $\sim$1$\times$10$^{-3}$ Torr and filled with Cl$_2$ gas at 0.5 Bar, then sealed.
Using an electric two zone tube furnace, one end of the quartz tube was heated up to 780 $^\circ$C while the other to 700 $^\circ$C, and the environment was maintained for 100 hours.
The obtained crystal size is typically $\sim$2$\times$2$\times$2 mm$^3$ with dark shinny facets.
The single phase of our samples was checked by X-ray diffraction.
The lattice parameters were obtained by analyzing the XRD patterns using FullProf\cite{fullprof}.

The x-ray absorption experiments were performed at the 2A elliptically polarized undulator beam line in the Pohang Light Source.
Fresh sample surfaces were obtained by scraping {\it in situ} with a diamond file at the pressure of $\sim$5$\times$10$^{-10}$ Torr.
The absorption spectra with the energy resolution of $\sim$0.2 eV were acquired at 120 K in a total electron yield mode using $\sim$98\% linearly polarized light, and were normalized to the incident photon flux.

The obtained XAS spectra were analyzed within the framework of the CICM with full multiplets\cite{vdLaan, deGroot, Cho, Noh1}.
In the pyrite nickel dichalcogenide (space group $Pa\overline{3}$) as shown in the right inset of Fig.\ref{fig1}, the cluster is a Ni[(S,Se)$_2$]$_6$ octahedron.
Six S/Se ligands form a regular octahedron and one Ni cation is located at the center of the octahedron.
Due to the relatively low electron affinity of the chalcogen ions, each ligand at the corners of the octahedron has a strong $\sigma$-bonding with another ligand, forming a dimer of (S,Se)$^{2-}_2$.
Divalent cation Ni$^{2+}$ is under the crystal electric field of $O_h$ symmetry, which induces a splitting of five 3$d$ orbitals into triply degenerate $t_{2g}$ and doubly degenerate $e_g$ states.
The splitting energy (10$Dq$) is set to 1.7 eV in this work.
The configurational interaction is characterized by the on-site $d$-$d$ Coulomb repulsion ($U\equiv E(d^{n+1})+E(d^{n-1})-2E(d^{n})$), the ligand $p$ to Ni 3$d$ charge transfer energy ($\Delta\equiv E(d^{n+1}\underline{L})-E(d^{n})$, $\underline{L}$ a ligand hole), and the hybridization strength between Ni 3$d$ and ligand $p$ states ($V\equiv \langle d|H|p\rangle$), which can be expressed with the Slater-Koster overlap integrals according to the point group symmetry of the cluster.
For the final state, the average Coulomb attraction between a 2$p$ core hole and a 3$d$ electron is set to 1.4$U$.
In this work, $U$ was set to 5.0 eV if not otherwise specified, and the others were used as fitting parameters.
The initial/final states are spanned over $|d^8\rangle$ $\oplus$ $|d^9\underline{L}\rangle$ / $|\underline{2p}d^9\rangle$ $\oplus$ $|\underline{2p}d^{10}\underline{L}\rangle$ configurations, respectively.
For full multiplet calculations of each configuration state, Slater-Condon parameters for Ni 3$d$-3$d$ and 2$p$-3$d$ ($F^2_{dd}$=12.234, $F^4_{dd}$=7.598, $F^2_{pd}$=7.721, $G^1_{pd}$=5.787, and $G^3_{pd}$=3.291 eV) are exploited with a reducing factor 0.8 to take the screening effect into account for the core hole.

\section{Results and Discussion}

\begin{figure}
\includegraphics[width=9cm]{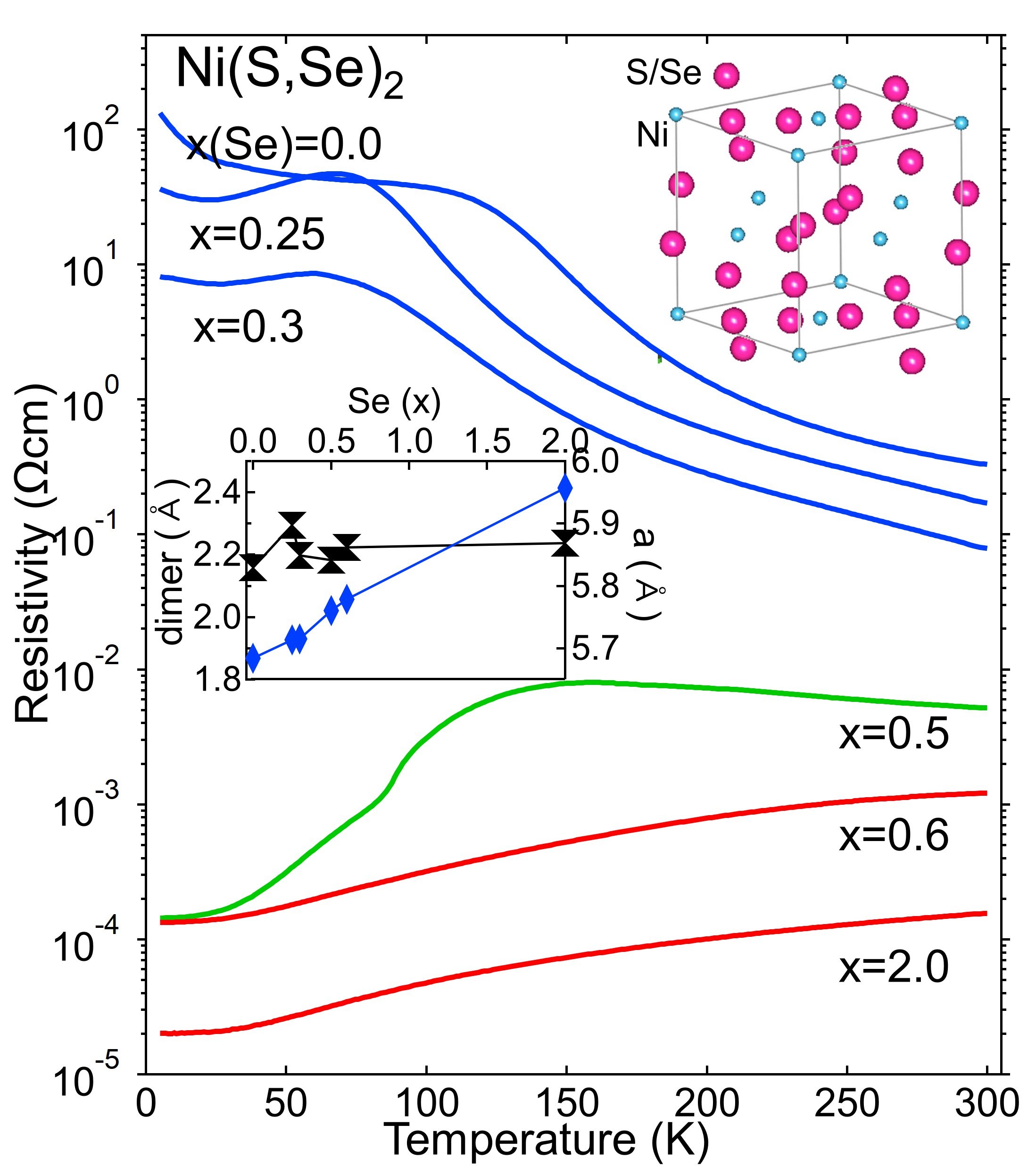}
\caption{Resistivity vs. temperature curves of NiS$_{2-x}$Se$_x$. Above $\sim$120 K, samples with $x$ $\leq$ 0.5 show the insulating behavior. (Left inset) Lattice constant $a$ (right axis, blue) / Ligand-ligand dimer distance (left axis, black) as a function of Se concentration $x$. (Right inset) Unit cell of the pyrite structure.}
\label{fig1}
\end{figure}

Figure \ref{fig1} shows the resistivity vs. temperature curves of our pyrite structure NiS$_{2-x}$Se$_x$ single-crystalline samples.
NiSe$_2$ and NiS$_{1.4}$Se$_{0.6}$ show a typical metallic behavior for a whole temperature range.
Meanwhile, NiS$_{1.5}$Se$_{0.5}$ show an MIT around 120 K, above which it becomes insulating.
As the Se concentration $x$ decreases to 0.25, the transition temperature goes down to $\sim$50 K.
At $x$=0, the slope of the resistivity curve changes around $\sim$120 K and $\sim$20 K, but the phase stays insulating.
These behaviors are very consistent with the previous observations on this system\cite{Bouchard}.
The left inset of Fig.~\ref{fig1} shows the lattice constant $a$ of the pyrite structure (right axis, blue) and the average distance of the ligand-ligand dimers (left axis, black) as a function of $x$, respectively\cite{Otero}.
The constants were extracted from a Rietveld analysis of the XRD patterns of the samples, showing a good agreement with the Vegard rule\cite{Vegard}.
The dimer distances show a little fluctuation, but no abrupt change is found.

\begin{figure}
\includegraphics[width=9cm]{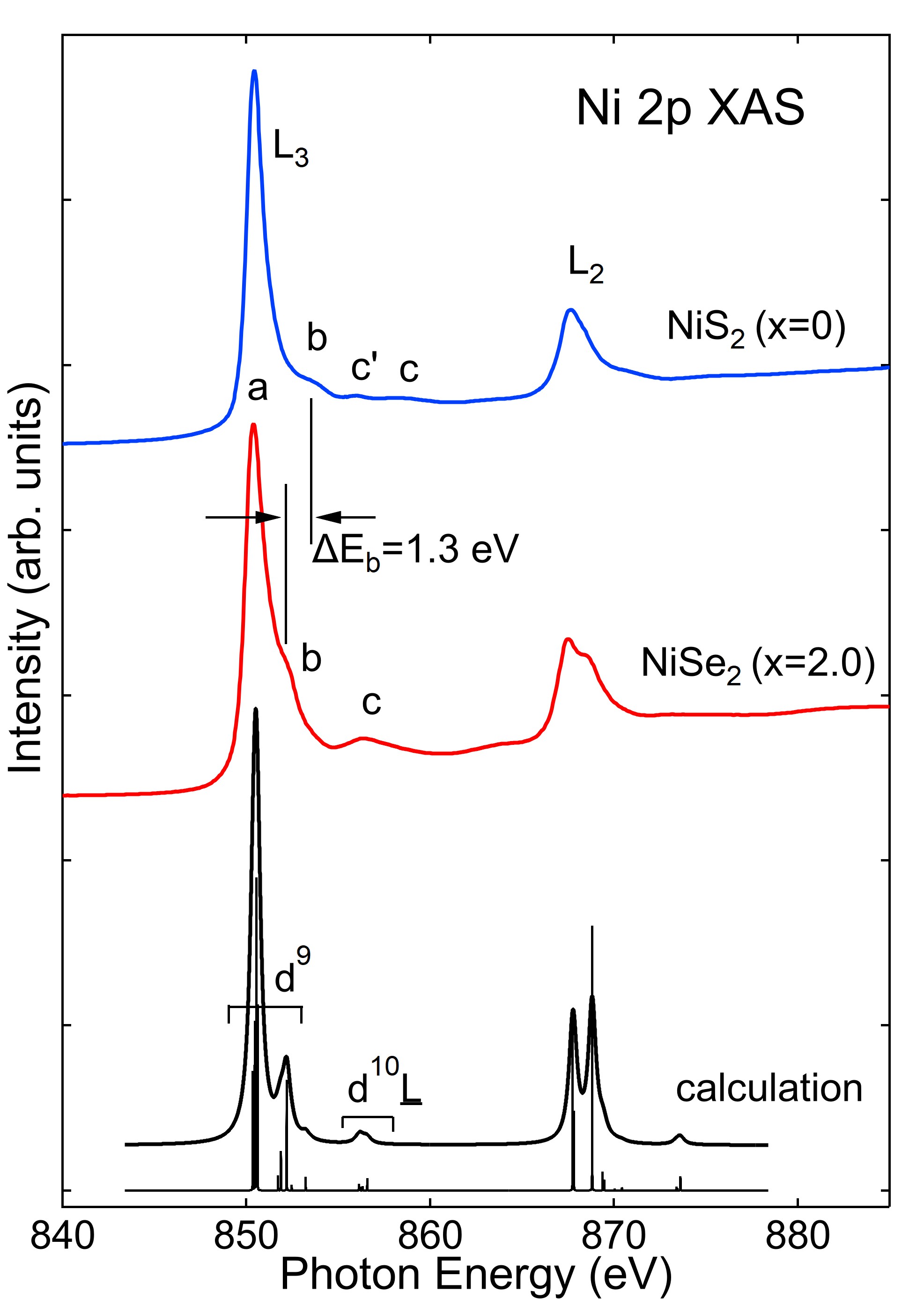}
\caption{\label{fig2}
Ni 2$p$ X-ray absorption spectra of NiS$_{2}$, NiSe$_{2}$, and a simulated spectrum based on the CICM with full multiplets.}
\end{figure}

The Ni 2$p$ XAS spectra of the end members ($x$=0, 2.0) is displayed  in Fig.~\ref{fig2}.
For comparison, a calculated spectrum based on the CICM with full multiplets ($\Delta$=3.5, $U$=5.0, $Q$=6.0, $V_{pd\sigma}$=1.2, 10$Dq$=1.0 eV) is shown together.
The major difference between the two end-members in the experimental data is the relative position ($\Delta{E}_b$$\sim$1.3 eV) of the $b$ peak in the $L_3$ edge.
In the insulating phase, the $b$ peak is more away from the main peak ($a$).
The physical analysis of the 2$p$ XAS of Ni$^{2+}$ ion under a crystal electric field of $O_h$ symmetry is well documented in Ref.\cite{vdLaan}.
The ground state symmetry is $^3A_{2g}$ ($e_g^2$) and the configurational character of $a$ and $b$ is $|\underline{2p}_{3/2}d^9\rangle$ for both cases, but the energy difference $\Delta{E}_b$ depends mainly on the $p$-$d$ charge transfer energy $\Delta$, and the $d$-$p$ hybridization strength $V$.
This is because the ground state is mixed with $|d^8\rangle$ and $|d^9\underline{L}\rangle$.
The calculated spectrum has all key features $a$, $b$, and $c$($c'$) to explain the experimental spectra.
It is well known that the Ni $L_2$ edge spectra are much more influenced by the ligand hole band width\cite{vdLaan}, so here we do not include $L_2$ edge fitting.

\begin{figure}
\includegraphics[width=9cm]{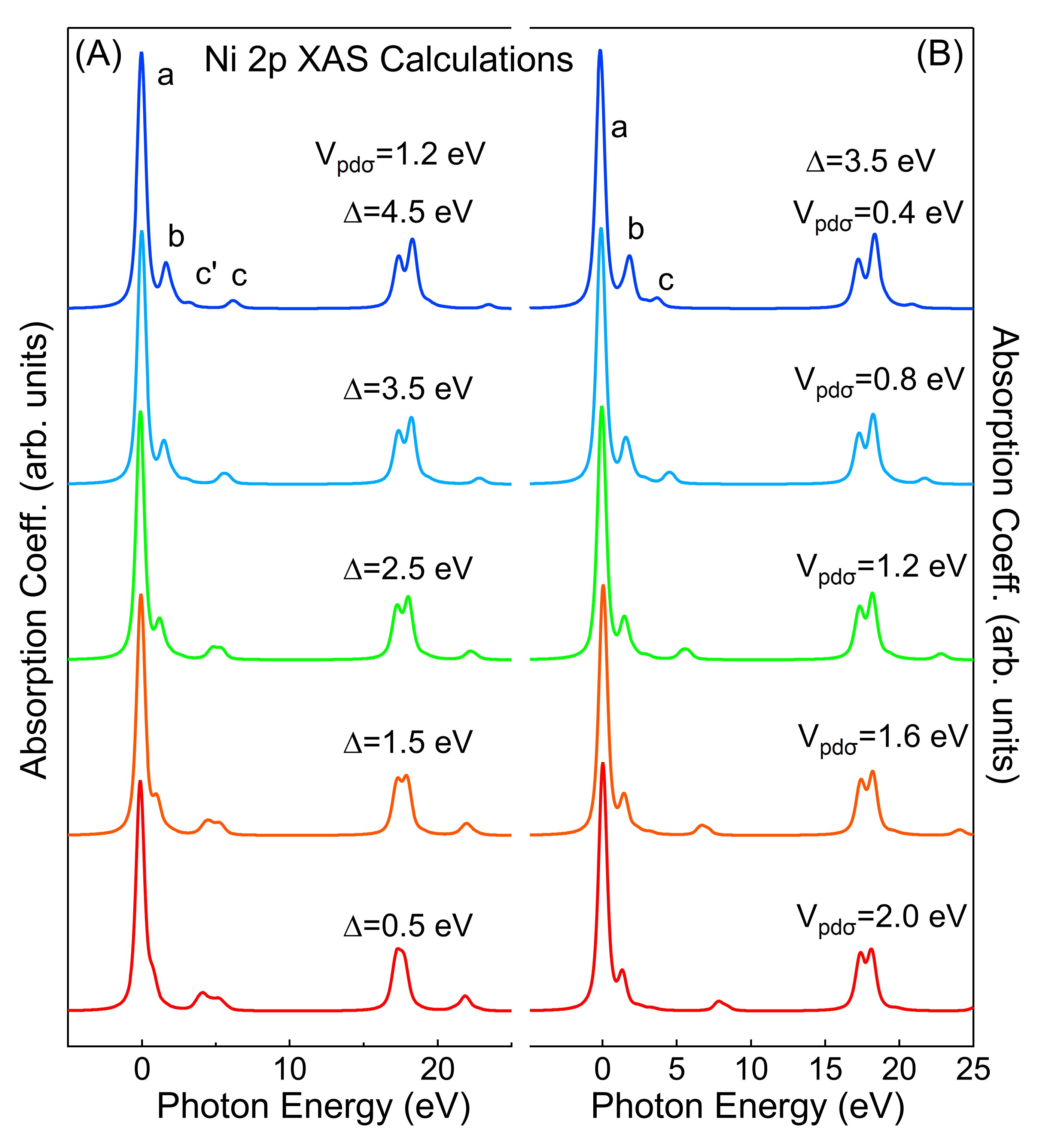}
\caption{\label{fig3}
Dependency of the calculated spectra on (A) ligand $p$-Ni 3$d$ charge transfer energy $\Delta$ and (B) Ni 3$d$-ligand $p$ hybridization strength $V_{pd\sigma}$.
Other parameter values are $U$=7, $Q$=8.75, and 10$Dq$=1.7 eV, respectively.
}
\end{figure}

In order to get more hints for fitting strategy, we first looked up the dependency of the calculated spectra on the four major characterizing parameters of the configurational interaction cluster model.
It is revealed that the $p$-$d$ charge transfer energy $\Delta$ and the $d$-$p$ hybridization strength $V$ are dominant on all split energies of the peaks, while the on-site Coulomb energy $U$ and the crystal electric field strength 10$Dq$ affect little on the energy $E_b$.
In Fig.~\ref{fig3}, we show how the calculated spectra are evolved with the two dominant parameters $\Delta$ and $V$.
In these simulations, the other parameters are set to $U$=7, $Q$=8.75, and 10$Dq$=1.7 eV, respectively.
As $\Delta$ decreases from 4.5 to 0.5 eV, all split energies get reduced as shown in Fig.~\ref{fig3}(A).
Meanwhile, as $V_{pd\sigma}$ decreases from 2.0 to 0.4 eV as shown in Fig.~\ref{fig3}(B), the energy difference ($E_{ab}$) between $a$ and $b$ increases slightly and the energy difference ($E_{bc}$) between $b$ and $c$ decreases drastically.
These behaviors suggest how the XAS spectra can be explained in terms of the CICM.
Se substitution in NiS$_2$ is expected to reduce the charge transfer energy and to increase the hybridization strength due to the smaller electro-negativity of Se and more extended nature of Se 4$p$ orbitals.
Even though these two parameters can affect on the split energy $\Delta{E}_{ab}$ in a similar way, they have a different effect on $\Delta{E}_{bc}$.
Thus, if we carefully observe the XAS spectral evolution as a function of Se concentration and compare them with the above parameter dependency, a decisive clue on the microscopic origin of the MIT in NiS$_{2-x}$Se$_x$ can be caught as follows.

\begin{figure}
\includegraphics[width=9cm]{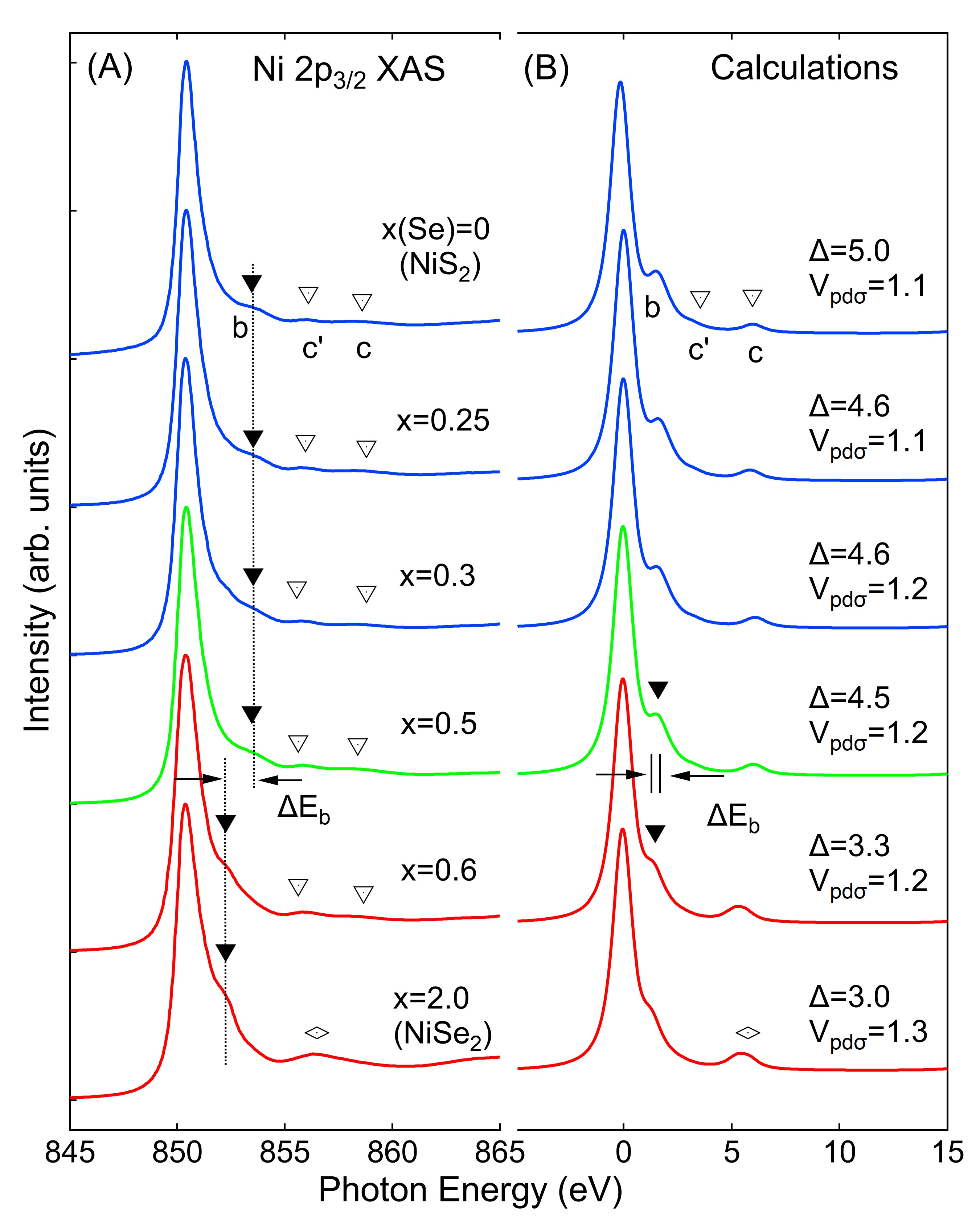}
\caption{\label{fig4}
(A) Experimental Ni 2$p_{3/2}$ XAS of NiS$_{2-x}$Se$_x$. (B) Fitted XAS based on the CICM calculation.}
\end{figure}

Figure~\ref{fig4}(A) shows our experimental XAS spectra at Ni $L_3$(2$p_{3/2}$) edge as a function of Se concentration.
Here, an interesting feature is found in the XAS spectra.
Although we obtained the spectra over the whole $x$ range, only two spectral shapes are found.
All insulating samples (0.0 $\leq$ $x$ $\leq$ 0.5) exhibit the $b$ peak at 853.6 eV, while all metallic samples (0.6 $\leq$ $x$ $\leq$ 2.0 ) do at 852.3 eV.
This dichotomic behavior is quite surprising when we consider the small difference of Se concentration between $x$=0.5 and 0.6 samples and the corresponding crystal structure difference.
Also, it is worth noting that the $c$/$c'$ peaks stay almost at the same energy position in spite of the wide Se concentration changes.
These two features in the spectral evolution strongly indicate the following two facts.
First, the Se substitution does not vary much the $d$-$p$ hybridization strength, contrary to our simple expectation.
Second, the $p$-$d$ charge transfer energy is the key control parameter in the MIT of this system, but the control mechanism is not a result of a simple average effect of the Se and S electro-negativities.

In Fig.~\ref{fig4}(B), our fitting results based on the CICM with full multiplets are displayed.
Here, we set to $U$=5.0, $Q$=7.0, and 10$Dq$=1.7 eV, respectively, and 0.9 eV Lorentzian broadening in full with at half maximum was applied to simulate the core hole life time effect.
As we approach the simulations from NiS$_2$ ($x$=0) to NiSe$_2$ ($x$=2.0), $\Delta$ decreases from 5.0 to 3.0 eV, and $V_{pd\sigma}$ a little increases from 1.1 to 1.3 eV.
A semi-quantitative spectral evolution is well reproduced in our fitting, and at least $\sim$1 eV abrupt change is essential in the $p$-$d$ charge transfer energy in order to simulate the MIT.
Note that $\Delta{E}_b$ in the simulation is much smaller than the experimental value.
The true change of the $\Delta$ value can be much larger than this semi-quantitative fitting result.

Our XAS spectra and the CICM analysis clarify that the MIT of NiS$_{2-x}$Se$_x$ is controlled by the ligand $p$-Ni 3$d$ charge transfer energy, but that the control mechanism is not a simple averaging effect of Se substitution into sulfur sites.
One possible scenario about the microscopic origin is a percolation of Ni[(S,Se)$_2$]$_6$ octahedra with at least one Se-Se dimer in the system.
As we checked the bonding distance of the dimers in the left inset of Fig.~\ref{fig1}, the distance itself does not show a clear correlation with the MIT although the fluctuation is relatively large.
However, it is noticeable that the dimer bonding state is the key factor to control the $\Delta$ value by a non-averaging way.
There are three kinds of dimers S-S, S-Se, and Se-Se in this system.
Considering the XAS spectra of the end members NiS$_2$ and NiSe$_2$, we can assume that only Ni[(S,Se)$_2$]$_6$ octahedra with at least one Se-Se dimer has a small enough $\Delta$ to induce a local screening effect, and that a long range connection of those octahedra makes a metallic transition of the system by a macroscopically transformed screening effect.
Then, this is a percolation problem of the Se-Se dimers in the pyrite structure.
In the structure, the dimers form a fcc structure (See Fig.\ref{fig1} right inset), and a local connection of the Ni[(S,Se)$_2$]$_6$ octahedra is equivalent to a site percolation of the dimers up to next nearest neighbor in the fcc structure.
Unfortunately, the exact percolation threshold probability $p_{th}$ to this problem is not found in literature, but by considering the similar cases, for example, simple cubic lattice up to next nearest neighbor (coordination number $z$=18)\cite{Kurzawski}, we can guess $p_{th}$$<$0.137.
This corresponds to the Se concentration $x<$0.740 under the assumption of random site occupation of Se atoms.
Although this is a little higher concentration than the experimental value $x_c$=0.55\cite{Bouchard}, major features of the MIT in this system are well explained in terms of Se-Se dimer percolation.

\section{Summary}

We synthesized the single crystalline NiS$_{2-x}$Se$_x$ samples with pyrite structure by the Cl$_2$ vapor transport method, analyzed the crystal structure to extract the lattice parameters, and measured the resistivity vs. temperature as a function of Se concentration, which show a good agreement with literature.
With the qualified samples, the electronic structure change has been studied by Ni $L$-edge XAS.
The XAS spectra show distinct features in Ni $L_3$ edge, indicating whether the system is insulating or metallic.
These features can be semi-quantitatively explained within the framework of the CICM.
In the S-rich region, relatively large charge-transfer energies ($\Delta \sim 5$ eV) from ligand $p$ to Ni 3$d$ states and a little small $p$-$d$ hybridization strength ($V_{pd\sigma} \sim 1.1$ eV) can reproduce the experimental spectra in the CICM calculation, and vice versa in the Se-rich region.
Our analysis result is consistent with the ZSA scheme that the systems in S-rich side ($x \leq$ 0.5) are a charge transfer insulator whose gap opening/closure depends on the relative size of the $p$-$d$ charge transfer energy ($\Delta$), the $d$-$d$ Coulomb energy ($U$), and the $p$-$d$ hybridization strength (V$_{pd\sigma}$), but it also requires an abrupt change of the $\Delta$ value, which cannot be a result of simple averaging effect of the S and Se electro-negativities.
As a possible microscopic origin, we propose a percolation scenario where a long range connection of Ni[(S,Se)$_2$]$_6$ octahedra with Se-Se dimers plays a key role to the gap closure through an abrupt decrease of $\Delta$.

\begin{acknowledgments}
This work was supported by the National Research Foundation (NRF) of Korea Grant funded by the Korean Government (Grant Nos. 2013R1A1A2058195 and 2016R1D1A3B03934980). The experiments at PLS were supported in part by MEST and POSTECH.
\end{acknowledgments}


\begin{references}
\bibitem{Bouchard} R. J. Bouchard, J. L. Gillson, and H. S. Jarrett, Mater. Res. Bull. \textbf{8}, 489 (1973).
\bibitem{Jarrett}  H. S. Jarrett \textit{et al.}. Mater. Res. Bull. \textbf{8}, 877 (1973).
\bibitem{Wilson} J. A. Wilson, The Metallic and Nonmetallic States of Matter, (Taylor \& Francis, London, 1985), p. 215.
\bibitem{Husmann} A. Hsmann, D S. Jin, Y. V. Zastavker, T. F. Rosenbaum, X. Yao, J. M. Honig, Science \textbf{274}, 1874 (1996).
\bibitem{ZSA} J. Zaanen, G. A. Sawatzky, and J. W. Allen, Phys. Rev. Lett. \textbf{55}, 418 (1985).
\bibitem{Kautz} R. L. Kautz, M. S. Dresselhaus, D. Adler, and A. Linz, Phys. Rev. B \textbf{6}, 2078 (1972).
\bibitem{Mabatah} A. K. Mabatah, E. J. Yoffa, P. C. Eklund, M. S. Dresselhaus, and D. Adler, Phys. Rev. B 21, 1676 (1980).
\bibitem{Matssura} A. Y. Matsuura, Z.-X. Shen, D. S. Dessau, C.-H. Park, T. Thio, J. W. Bennett, O. Jepsen, Phys. Rev. B 53, R7584 (1996).
\bibitem{Mamiya} K. Mamiya, T. Mizokawa, A. Fujimori, Phys. Rev. B \textbf{58}, 9611 (1998).
\bibitem{Perucchi} A. Perucchi, C. Marini, M. Valentini, P. Postorino, R. Sopracase, P. Dore, P. Hansmann, O. Jepsen, G. Sangiovanni, A. Toschi, K. Held, D. Topwal, D. D. Sarma, and S. Lupi, Phys. Rev. B \textbf{80}, 073101 (2009).
\bibitem{ARPES} A. Damascelli,
%    Probing the Electronic Structure of Complex Systems by ARPES.
    Physica Scripta \textbf{T109}, 61 (2004).
\bibitem{DMFT} A. Georges, G. Kotliar, W. Krauth, and M. J. Rozenberg, Rev. Mod. Phys. \textbf{68}, 13 (1996).
\bibitem{Jeon} A. J. Kim, M. Y. Choi, and G. S. Jeon, J. Kor. Phys. Soc. \textbf{64}, 268 (2014).
\bibitem{Kunes} J. Kune\v{s}, L. Baldassarre, B. Sch¡§achner, K. Rabia, C. A. Kuntscher, Dm. M. Korotin, V. I. Anisimov, J. A. McLeod, E. Z.Kurmaev, and A. Moewes, Phys. Rev. B \textbf{81}, 035122 (2010).
\bibitem{Moon} C.-Y. Moon, H. Kang, B. G. Jang, and J. H. Shim, Phys. Rev. B \textbf{92}, 235130 (2015).
\bibitem{Xu} H. C. Xu, Y. Zhang, M. Xu, R. Peng, X. P. Shen, V. N. Strocov, M. Shi, M. Kobayashi, T. Schmitt, B. P. Xie, and D. L. Feng, Phys. Rev. Lett. \textbf{112}, 087603 (2014).
\bibitem{Charnock} J. M. Charnock, C. M. B. Henderson, J. F. W. Mosselmans, and R. A. D. Pattrick, Phys. Chem. Minerals \emph{23}, 403 (1996).
\bibitem{Suga} S. Suga, A. Kimura, T. Matsushita, A. Sekiyama, S. Imada, K. Mamiya, A. Fujimori, H. Takahashi, and N. Mori, Phys. Rev. B \emph{60}, 5049 (1999).
\bibitem{fullprof} T. Roisnel and J. Rodriguez-Carvajal, FullProf, 2000.
\bibitem{vdLaan} G. van der Laan, J. Zaanen, G. A. Sawatzky, R. Karnatak, and J.-M. Esteva, Phys. Rev. B 33, 4253 (1986).
\bibitem{deGroot} F. M. F. de Groot and J. C. Fuggle, Phys. Rev. B \textbf{42}, 5459 (1990).
\bibitem{Cho} K. Cho, H. Koh, J. Park, S.-J. Oh, H.-D. Kim, M. Han, J.-H. Park, C. T. Chen, Y. D. Kim, J.-S. Kim, and B. T. Jonker, Phys. Rev. B 63, 155203 (2001).
\bibitem{Noh1} H.-J. Noh, S. Yeo, J.-S. Kang, C. L. Zhang, S.-W. Cheong, S.-J. Oh, P. D. Johnson, Appl. Phys. Lett. \textbf{88}, 081911 (2006).
\bibitem{Otero} R. Otero, J. L. Martin de Vidales, and C. de las Heras, J. Phys. Condens. Matter \textbf{10} 6919 (1998).
\bibitem{Vegard} L. Vegard, Zeitschrift f\"{u}r Physik \textbf{5} 17 (1921).
\bibitem{Kurzawski} L. Kurzawski and K. Malarz, Reports on Math. Phys. \textbf{70}, 163 (2012).


\end{references}
\end{document}